\documentclass[traditabstract]{aa} 

\pdfoutput=1 

\usepackage{graphicx}
\usepackage{txfonts}

\def\0{\phantom0}

\def\kms{km s$^{-1}$}

\def\obj{SDSS~J0013+1523}

\def\OIII{[\ion{O}{III}]}
\def\OII{[\ion{O}{II}]}

\begin{document}
\title{First case of strong gravitational lensing by a QSO : SDSS~J0013+1523 at z = 0.120{\thanks{
Some of the data  presented herein were obtained at the W.M. Keck Observatory, which is  
operated as a scientific partnership among the California Institute of  
Technology, the University of California and the National Aeronautics  
and Space Administration. The Observatory was made possible by the  
generous financial support of the W.M. Keck Foundation. This program also 
makes use of the data collected by the SDSS collaboration and released in DR7.}}}


\titlerunning{Strong gravitational lensing by a QSO}

\author{}

\author{F.~Courbin\inst{1} \and M.~Tewes\inst{1} \and S.G. Djorgovski\inst{2} \and D.~Sluse\inst{3}\thanks{Alexander Von Humboldt fellow} \and A.~Mahabal\inst{2} \and F.~R\'erat\inst{1} \and G.~Meylan\inst{1} }

   \institute{Laboratoire d'astrophysique, Ecole Polytechnique
     F\'ed\'erale de Lausanne (EPFL), Observatoire de Sauverny,
     CH-1290 Versoix, Switzerland
     \and
     Division of Physics, Mathematics, and Astronomy, California Institute of Technology, Pasadena, CA 
     91125, USA
     \and     
     Astronomisches Rechen-Institut 
    am Zentrum f\"ur Astronomie der Universit\"at Heidelberg,
    M\"onchhofstrasse 12-14, D-69120 Heidelberg, Germany
  } 

   \date{Received; accepted }

   \abstract{We present the first case of strong gravitational lensing by a QSO : \obj
   at $z=0.120$. The discovery is the result of a systematic search for emission lines redshifted behind
   QSOs, among 22 298 spectra of the SDSS data release~7. Apart from the $z = 0.120$ 
   spectral features of the foreground QSO, the spectrum of \obj\, also displays 
   the \OII\ and H$\beta$ emission lines and the \OIII\ doublet, all at the same redshift, $z = 0.640$. 
   Using sharp Keck adaptive optics K-band images obtained using laser guide stars, we unveil two objects 
   within a radius of 2\arcsec\ from the QSO. Deep Keck optical spectroscopy clearly confirms one of these 
   objects at $z = 0.640$ and shows traces of the \OIII\, emission line of the second object, also at $z = 0.640$.
   Lens modeling suggests that they represent two images of the same $z = 0.640$
   emission-line galaxy. Our Keck spectra also allow us to measure the redshift of an intervening galaxy at 
   $z = 0.394$, located 3.2\arcsec\, away from the line of sight to the QSO.
   If the $z = 0.120$ QSO host galaxy is modeled as a singular isothermal sphere, its mass within the 
   Einstein radius is $M_E(r<1\, h^{-1}\,kpc) = 2.16 \times10^{10}\ h^{-1}\ M_{\odot}$ and its velocity dispersion
   is $\sigma_{\rm SIS} = 169$ \kms. This is about 1-$\sigma$ away from the velocity dispersion estimated from the 
   width of the QSO H$\beta$ emission line, $\sigma_*(M_{BH})= 124 \pm 47$ \kms.
   Deep optical HST imaging will be necessary to constrain 
   the total radial mass profile of the QSO host galaxy using the detailed shape of the lensed source. 
   This first case of a QSO acting as a strong lens on a more distant object opens new directions 
   in the study of QSO host galaxies.} 

   \keywords{Gravitational lensing: QSO -- QSOs: individual (\obj) -- QSOs: host galaxies.}

   \maketitle
%

\section{Introduction}

With the growing number of ongoing and future wide-field surveys, strong gravitational lensing is becoming a powerful tool to weigh individual galaxies and study their total radial mass profiles. Historically, strong lenses have long been source-selected, i.e., they were found by identifying gravitationally lensed multiple images of distant sources such as QSOs, either in the optical (e.g., SDSS; Oguri et al. \cite{OGU06}) or in the radio (e.g., the CLASS survey; e.g., Browne et al.~\cite{BROWN03}, Myers et al.~\cite{MYERS03}). The disco\-veries of multiply imaged QSOs, by numerous independent teams over two decades, have led to a sample of about 100 strongly lensed QSOs. Detailed HST imaging is now available for most of these objects mainly thanks to the CfA-Arizona Space Telescope LEns Survey (CASTLES; Mu{\~n}oz et al.\cite{MUN98}, Impey et al. \cite{IMP98}). The second largest sample of source-selected systems is the one obtained in the COSMOS field (Faure et al. \cite{FAU}), where the sources are almost always emission-line galaxies. 

Source-selected samples tend to yield lensing galaxies spanning a broad range of physical properties, i.e., effective radii, Einstein radii, dark matter profiles, total masses, and environments. With current wide-field surveys such as the SDSS, it is now also possible to build lens-selected samples, where the lenses can be chosen with well-defined photometric properties. The largest sample of these lens-selected systems to date is the SLACS sample (e.g., Bolton et al. \cite{BOL}), where the lenses are color-selected early-type galaxies with $0.1 < z < 0.6$ and where the sources are emission line galaxies with $z>0.5$. While source-selected samples use optical or radio data to look for multiple images, SLACS uses SDSS optical spectra to look for emission lines redshifted behind the selected lensing galaxies, following the technique of Warren et al. (\cite{WAR}).

\begin{figure*}[t!]
\begin{center}
\includegraphics[width=18.cm, height=9.cm]{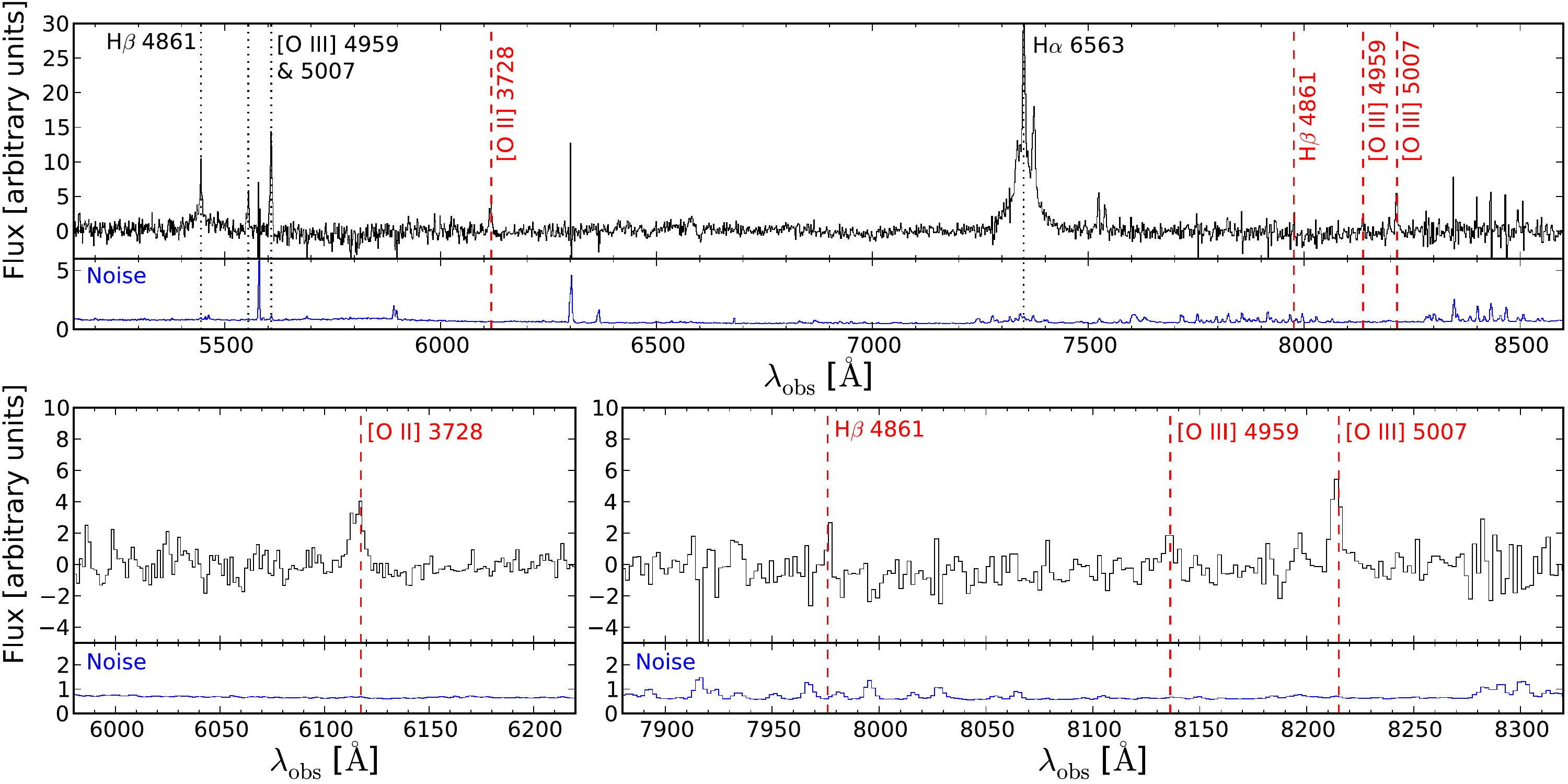}

\caption{{\it Top:} SDSS spectrum of \obj, after subtraction of the continuum, as provided by the SDSS pipeline. The emission lines associated with the foreground $z=0.120$ QSO are shown with vertical dotted lines (black) and the emission lines associated with the $z=0.640$ background object are shown with dashed lines (red). The noise spectrum is also displayed indicating the position of the strongest sky lines. {\it Bottom:} from left to right, zooms on the $z=0.640$ \OII, H$\beta$ and on the \OIII\, doublet. The noise spectrum does not show any strong feature at the position of the background emission lines. Note as well that the \OII\, doublet is spectrally resolved. No smoothing has been applied.}
\label{fig:SDSS_spectra}
\end{center}
\end{figure*}

Following the same approach, we started to compile a lens-selected sample where the foreground lenses are QSOs in place of early-type galaxies. Our long-term goal is to provide a direct measure of the total mass of QSO host galaxies and to use SLACS as a non-QSO ``control'' sample. Because the lensing QSOs are at low redshift ($z<0.4$), the range of follow-up applications is very rich, as their H$\beta$, H$\alpha$, \OIII\, emission lines are clearly measurable in the SDSS spectra. In addition, the angular size and luminosity of their host galaxies make it possible to carry out direct spectroscopy of the host (e.g.,  Lewate et al.~\cite{LETAW07}, Letawe et al.~\cite{LETAW04}, Courbin et al.~\cite{COU02}). Our sample will therefore allow us to test directly the scaling laws established between the properties of QSO emission lines, the mass of the central black hole, and the total mass of the host galaxies (e.g., Kaspi et al. \cite{KAS}, Bonning et al. \cite{BON}, Shen et al. \cite{SHEN}). 

In this paper, we focus on the discovery of the first case of a QSO, \obj\, (RA(2000): 00h 13m 40.21s; DEC(2000): +15$^{\circ}$ 23\arcmin\  12.1\arcsec; $z = 0.120$; $r=18.04$), producing two images of a background galaxy ($z = 0.640$) due to strong gravitational lensing.

\begin{figure}[h!]
\begin{center}
\includegraphics[width=8.9cm, height=8.7cm]{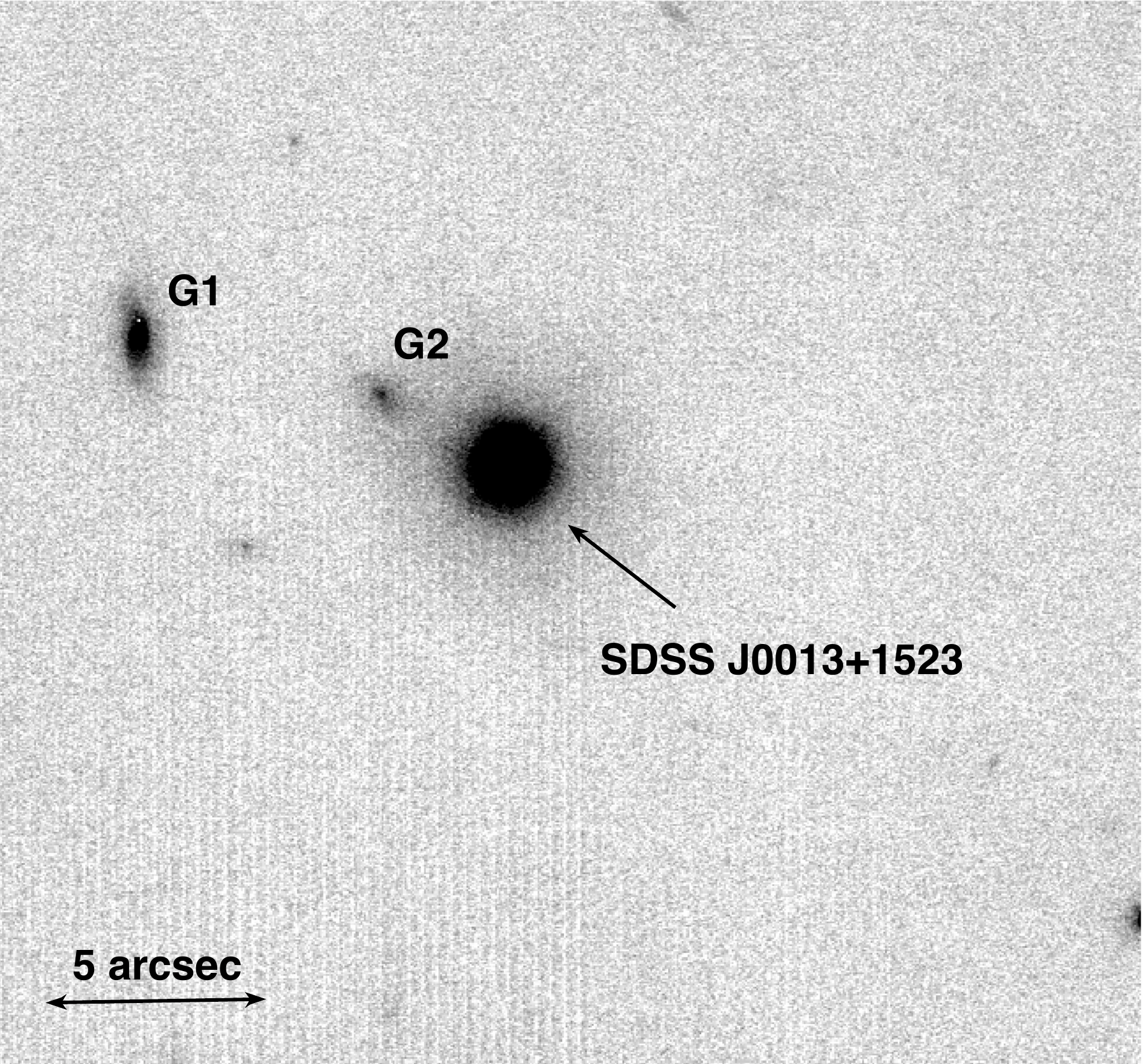}
\caption{ Full field of view of the Keck NIRC2+AO observations. No star is available to construct the PSF. Also indicated are galaxies G1 and G2, which may contribute to the external shear of the lens models. From our Keck spectra, galaxy G2 is at $z = 0.394$. North is to the top, east left.}
\label{fig:fullfield}
\end{center}
\end{figure}

\section{Spectroscopic search in SDSS}

We used the SDSS DR7 catalogue (Abazajian et al.~\cite{ABA09}) of spectroscopically confirmed QSOs 
 to build a sample of potential lenses, without applying any magnitude cut. We considered all QSOs in the catalogue with $0 < z_l < 0.7$, providing a sample of 22 298 potential lenses with SDSS spectra. In these spectra, we look for emission lines redshifted beyond the redshift of the QSO. Since the SDSS fiber is comparable in size  (3\arcsec\, in diameter) to the typical Einstein radius of our QSOs, the object responsible for the background emission lines is very likely to 
be strongly lensed (e.g., Bolton et al.~\cite{BOL}).

\begin{figure*}[t!]
\begin{center}

\includegraphics[width=18.4cm, height=6.1cm]{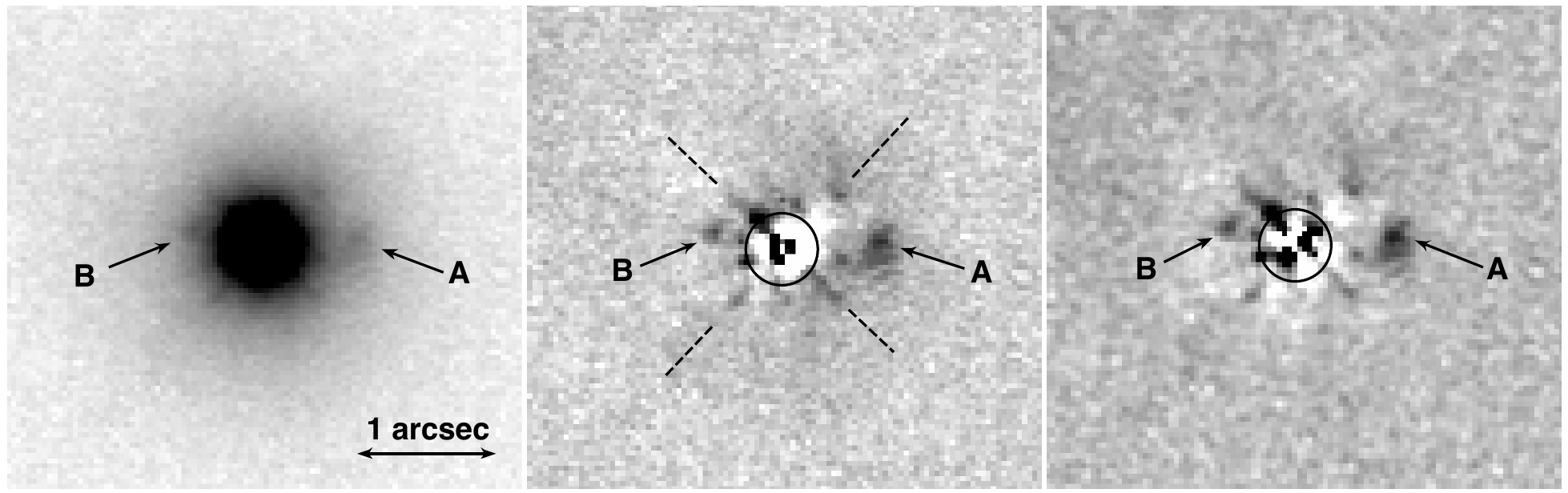}
\caption{{\it Left:} zoom on the central part of Fig.~\ref{fig:fullfield}. The resolution is 0.1\arcsec\, FWHM and the pixel scale is 0.0397\arcsec. Two objects, labelled A and B, appear in the vicinity of the QSO. {\it Middle:} residual after subtraction of the GALFIT model (Moffat+exponential disk), where objects A and B are now well visible. The circle indicates the position of the QSO. Its size, 0.5\arcsec\, in diameter, corresponds to the region where the subtraction may not be reliable, mainly due to the presence of the complex first Airy ring of the PSF. Spike-like features in the PSF are indicated by the dashed 
lines. {\it Right:} as in the middle panel, but a symmetrized PSF has been subtracted. This removes more effectively the Airy ring than an analytical model of the PSF.}
\label{fig:PSFsubtract}
\end{center}
\end{figure*}

Looking for emission lines in the background of QSOs is far more difficult than behind galaxies, for three reasons: (i) the QSO outshines the lensed background images, making follow-up imaging very challenging and (ii) in contrast to early-type galaxies, QSO spectra span a huge range of properties, both in the continuum and emission lines. It is therefore impossible to subtract a QSO template from the SDSS spectra and to search for emission lines in the residual spectrum. Finally, (iii), QSO spectra have many weak iron features that can mimic faint background emission lines. These lines lead to false positives that can be removed statistically thanks to
the huge number of QSO spectra available in SDSS.

Although our search technique will be the topic of a future paper, the main components of our strategy are the following. First we subtract the continuum of the QSO spectra using a spline fit. Then, we cross-correlate the continuum-subtracted spectra with several analytical templates of emission lines. We typically use 3 templates, one with only the optical H$\beta$ and \OIII\, doublet, a second to which we add the \OII\, doublet, and a third that also contains H$\alpha$. The cross-correlation provides us with a catalogue of candidates, a candidate being defined as a significant peak in the cross-correlation function of redshift higher than the foreground QSO. Each QSO spectrum has many possible candidates, of which we retain the 5 most significant. For those, we fit Gaussian profiles to each of the candidate emission lines and compute a quality factor, based on the total signal-to-noise ratio (S/N) in the spectroscopic lines. 

The quality factor is used to select the most likely candidates in terms of S/N. However, at this point, the ca\-ta\-logue of candidates still needs to be cleaned from false positives produced by residuals caused by either sky subtraction or faint features in the spectrum of the foreground QSO. Thanks to the large number of QSOs in the SDSS sample, these can be identified statistically and rejected efficiently. As a final step, we carry out a visual inspection of the most probable candidates. This process leaves us with about 10 candidates that have at least 3 significant emission lines and an additional 4 candidates with 4 emission lines. \obj\, is one of the latter. Its SDSS spectrum and the background emission lines 
at $z_s = 0.640$ are shown in Fig.~\ref{fig:SDSS_spectra}. 

If all 14 objects are actually lensed, the probability of finding these objects is 
$14 / 22,298 = 6.2 \times 10^{-4}$. 
This is about 8 times less than in SLACS, where the discovery rate is 0.005 (e.g., Auger et al.~\cite{AUGER09}). However,
our selection function is very different from SLACS, because of 
(i) the contrast bet\-ween the QSO and the background emission
lines, (ii) the regions of the QSO spectra that must be masked due to strong QSO spectral features, (iii) the requirement
to see multiple background emission lines simultaneously outside the masked regions, (iv) differences between the redshift distributions of QSOs in our sample and the galaxies in SLACS. A detailed study of the selection function (e.g., Dobler et al.~\cite{DOB08}) beyond the scope of the present discovery paper. However, if real, the difference between the SLACS discovery rate and ours may reflect a genuine difference between the lensing cross-sections of the two samples, 
i.e., between the physical properties of QSO and non-QSO galaxies.

\begin{figure*}[t!]
\begin{center}
\includegraphics[width=18.0cm, height=9cm]{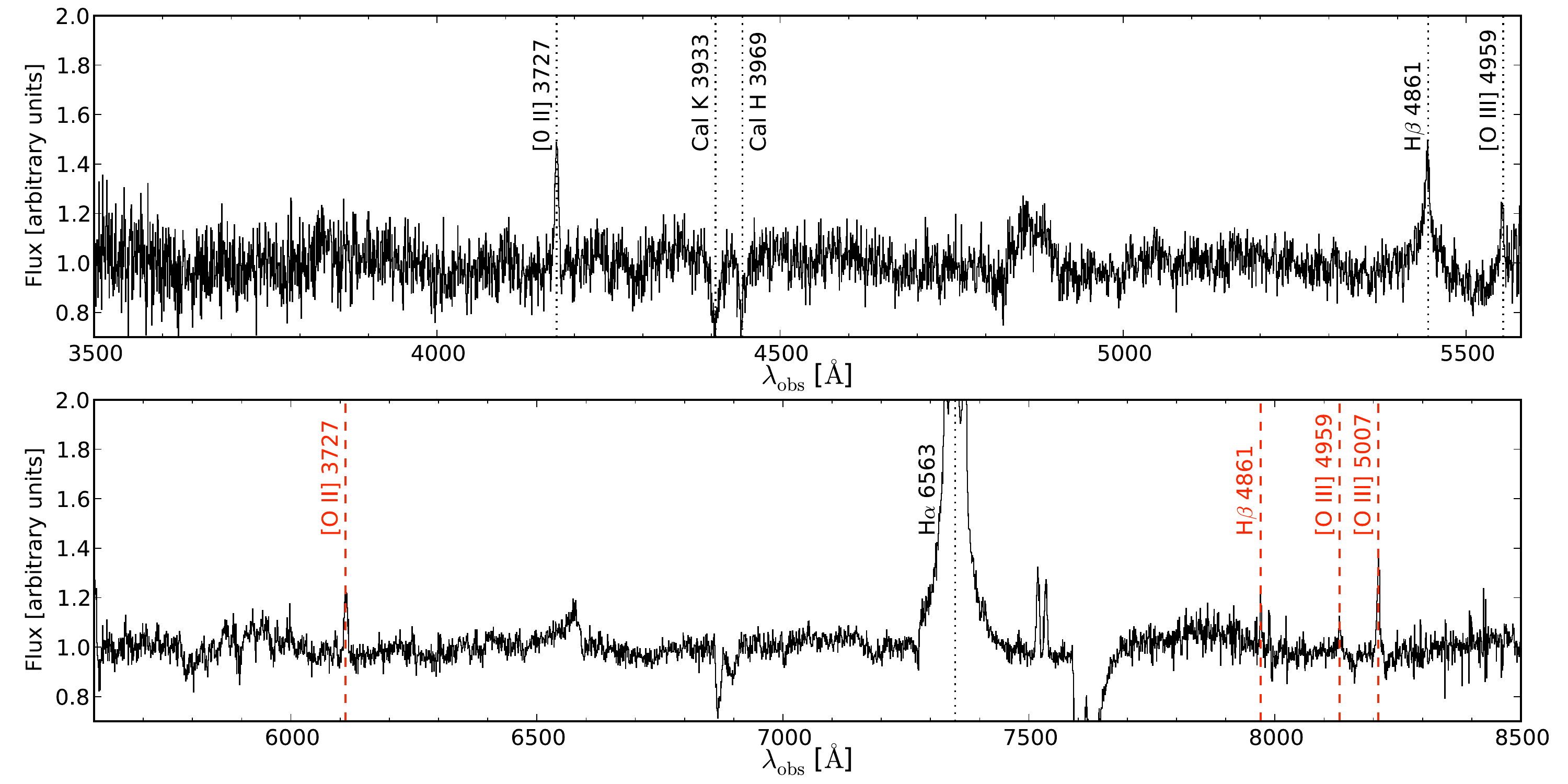}
\caption{Keck/LRIS spectrum of \obj, normalized to the continuum. 
As in Fig.~\ref{fig:SDSS_spectra}, the labels in black show the spectral
features associated with the foreground $z = 0.120$ QSO and the labels in red show the emission lines of the
$z = 0.640$ background galaxy. No smoothing has been applied.}
\label{fig:Keck_1Dspec}
\end{center}
\end{figure*}

\section{Keck adaptive optics imaging}

The imaging observations were obtained on the night of 11 September 2009 UT, using the NIRC2 instrument\footnote{\tt http://www2.keck.hawaii.edu/inst/nirc2/} in the laser guide star adaptive optics mode (LGSAO; Wizinowitch et al. 2006) at the 10-m Keck-2 telescope on Mauna Kea, Hawaii, in variable conditions.  Images  
were obtained in the $K^\prime$ band, using the ``wide field" NIRC2 camera with the FOV $\approx$ 40 arcsec, and the pixel scale of  0.0397\arcsec.  The observations consist of 42 dithered exposures of 40 seconds each. 
The data were reduced using standard procedures, leading to the combined image shown in Fig.~\ref{fig:fullfield}.

Owing to the small field of view of the NIRC2+AO imager, no star is available to construct a numerical 
point spread function (PSF), hence making image deconvolution impossible. We never\-theless subtracted the low
spatial frequencies of the PSF in two different ways.

First, we used the GALFIT package (Peng \cite{PEN02}) to fit analytical profiles to the data.  We tried a number of different models where the QSO is represented with either a Gaussian or a Moffat profile and where the host galaxy is represented by an elliptical Sersic, a de Vaucouleurs, or an exponential disk profile. We found that the best fits were thoses with a Moffat for the QSO and an exponential disk for the host. This model was subtracted from the 42 individual images and the difference images were then stacked and cleaned for cosmic-rays using sigma-clipping.

As a second method, we computed an estimate of the PSF directly from the QSO itself. The goal here was to remove the low frequency signal of the PSF (plus QSO host galaxy), by symmetrizing the image of the QSO.  For each of the 42 frames, we produced 10 different duplicates rotated by 36 degrees about the QSO centroid. We then took the median of these 10 rotated frames, hence producing a circularly symmetric PSF. Thanks to the median average, all small details in the PSF were removed, i.e., the spikes, any blob in the PSF, and the putative lensed images of a background object. The 42 PSFs were then subtracted from the data and the residual images were stacked together. 

\begin{table}[t!] 
  \caption{Summary of the Keck astrometry. The error bars in the positions are about 20 mas, i.e., about half a pixel.}
  \vspace{0.2cm}
  \begin{tabular}{lccc}
    \hline
    \hline
    Object & $x$(\arcsec) & $y$(\arcsec) &  Flux \\
    \hline
    QSO    &   $+$0.000         &    $+$0.000        &    $-$ \\
    A          &   $+$0.731         &    $+$0.027         &    1.00   \\ 
    B          &  $-$0.539        &     $+$0.132        &    0.72   \\     
    G1       &  $-$8.318         &    $+$2.873         & $-$ \\
    G2       &  $-$2.842         &    $+$1.610         & $-$ \\
    \hline
  \end{tabular}
  \label{tab:summary}
\end{table}

The results of these two ways of subtracting the QSO are shown in Fig.~\ref{fig:PSFsubtract}. The subtraction is found satisfactory for distances larger than 0.25\arcsec\ away from the QSO. In this area, the structures in the PSF-subtracted data change depending on the model we adopt in GALFIT. This is because the first diffraction ring of the PSF that cannot be modeled analytically. This ring is more accurately modeled by our second subtraction method, where we symmetrized the PSF. 

 Two obvious objects are found to the east and west of the QSO, labelled A and B in Fig.~\ref{fig:PSFsubtract}. These are outside the area where the PSF subtraction is uncertain and they are not located on the PSF spikes indicated by dashed lines in the middle panel of Fig.~\ref{fig:PSFsubtract}.  
We consider them as the best candidates for lensed images of a source at $z_s = 0.640$. All other
faint structures are either superimposed with PSF features or are not fitted by lens models (Sect.~\ref{models}). A summary of the relative astrometry and the flux ratio of objects A to B is given in Table~\ref{tab:summary}.

\section{Keck optical spectroscopy}

Deep Keck optical spectra were obtained on the night of 20 December 2009, using the 
LRIS\footnote{\tt http://www2.keck.hawaii.edu/inst/lris/} instrument (Oke et al. \cite{OKE95}). The
weather conditions were variable, with patchy clouds. The seeing at the moment
of the observations was 1.5\arcsec. 

A long slit was used (1.5\arcsec\ wide), with a position angle (PA) of 70$^\circ$, ensuring that we could observe
simultaneously the QSO as well as object A and galaxy G2. Object B also 
enters partly in the slit. The data consist of 3 exposures of 600 seconds each, with nodding along the slit
to minimize the effect of CCD traps. The wavelength range from 3500 \AA\, to 8500 \AA\, is 
covered using the dichroic \#560 that splits the beam into a blue and a red channels. The spectrum in
the blue channel is obtained using a 400 lines/millimeter grism and the spectrum in the red channel
is obtained with a 600 lines/millimeter grism.

The data were reduced using the IRAF\footnote{\tt http://iraf.noao.edu/} package, to perform the wavelength calibration, the 
sky subtraction, and the spectra combination in two dimensions. The pixel size in the blue channel 
is 0.6\AA\, in the spectral direction and 0.135\arcsec\, in the spatial direction. In the red channel the
spectral scale is 0.8 \AA\, per pixel and the spatial scale is 0.210\arcsec\, per pixel.
The final 1D spectrum was extracted, confirming all of
the $z=0.640$ emission lines seen in the SDSS spectrum  (Fig.~\ref{fig:Keck_1Dspec}). 
In addition to the QSO spectrum we
also obtained the spectrum of galaxy G2, allowing us to measure its redshift from the \OII, H$\beta$, and
\OIII\, emission lines, $z=0.394$.  This rules out the possibility 
that G2 is a counter image of object A. We do not detect the continuum of G2.

\begin{figure}[t!]
\begin{center}
\includegraphics[width=6.0cm, width=8.9cm]{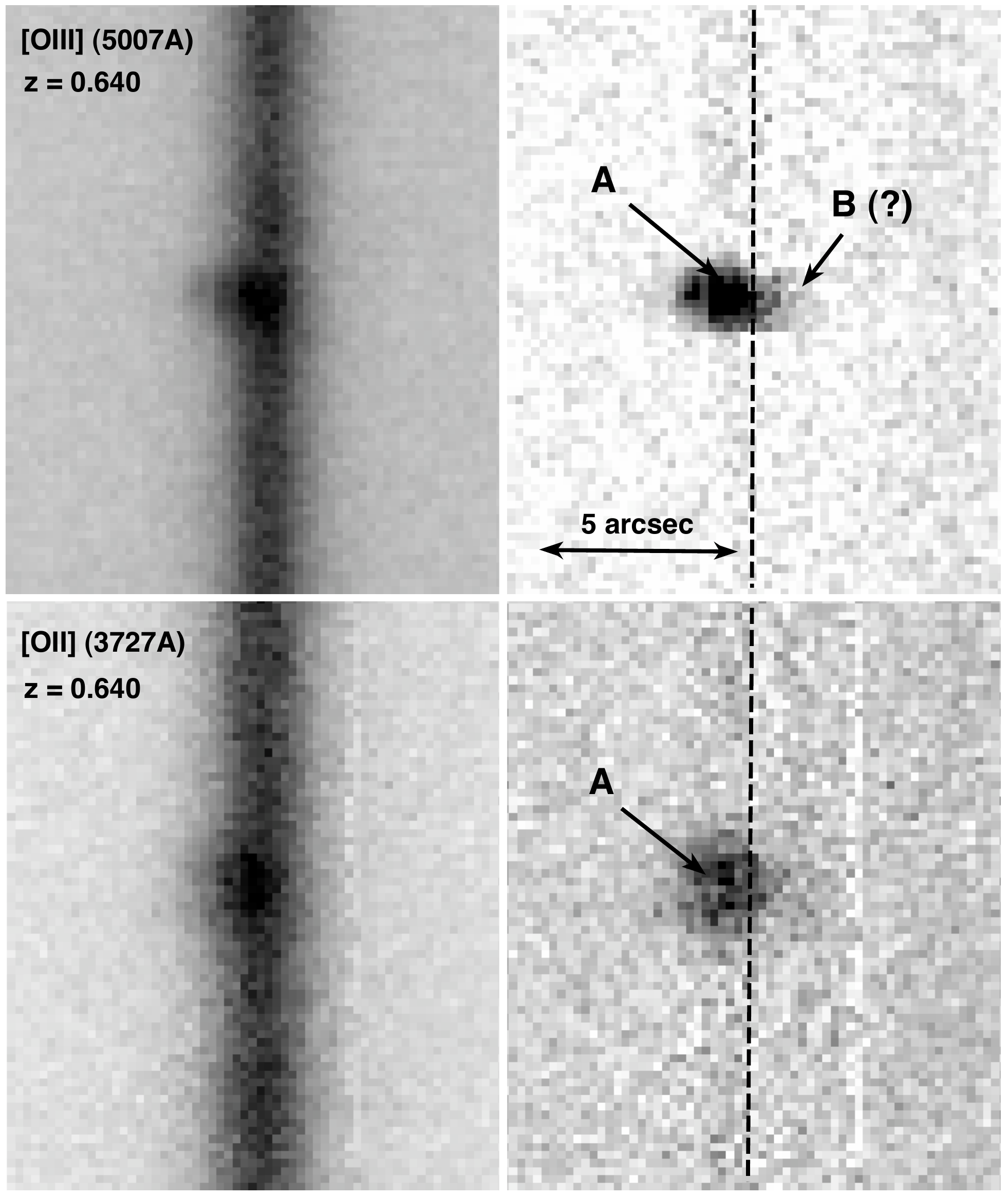}
\caption{Two-dimensional spectrum of \obj\, obtained with the Keck and LRIS. The spatial direction
runs along the x-axis and the spectral direction along the y-axis (red to the top). {\it Top:} zoom on the 
\OIII\ (5007\AA) emission line, redshifted to $z=0.640$. The left panel shows the original spectrum, while the
right panel shows the data after subtraction of the QSO spatial profile, centered on the dashed line. 
Object A is clearly detected, off-center with respect to the QSO, and traces of B may be present
at the position expected from the AO images. {\it Bottom:} same as above but for the \OII\ (3727 \AA) doublet.}
\label{fig:Keck_2Dspec}
\end{center}
\end{figure}

The seeing for the 2D spectra was poor, but sufficient to separate spatially the spectrum of the QSO from that of 
object A. This was achieved by subtracting a two-dimensional Moffat profile from the data, 
clearly unveiling the four $z=0.640$ emission lines seen in the 
1D spectrum, off-centered by 0.8\arcsec\, from the centroid of the QSO. This is almost exactly
the separation betwen the QSO and object A. We show in Fig.~\ref{fig:Keck_2Dspec} the result of the
QSO subtraction for the two brightest emission lines. Although the signal is very weak, we can
also detect a weak  \OIII\, emission line at the position expected for object B. Deeper observations with 
better seeing would be needed to provide firm conclusions.

\section{Lens modeling}
\label{models}

We show that the observed image
configuration is consistent with the gravitational lens hypothesis.
We model the mass distribution of the lens with a singular isothermal
sphere (SIS) and include the effect of the environment as an
external shear. This simple model has three parameters: the angular Einstein
radius $\theta_E$ of the SIS, the amplitude $\gamma$, and the position
angle $\theta_{\gamma}$ of the shear. The observational quantities
used to constrain the model are the relative positions of the lensed
images A and B and the flux ratio of these images. Since the
flux ratio can be affected by differential extinction, PSF
uncertainties, or microlensing, we assume an error of up to 20\% in the flux
ratio reported in Table~\ref{tab:summary}. The center of the SIS is
assumed to be coincident with the position of the lensing QSO. Using
the \texttt{lensmodel} lensing code developed by Keeton (2001), 
we find that the observed system can be perfectly
reproduced ($\chi^2 \sim$ 0) with the following model parameters
($\theta_E$, $\gamma$, $\theta_{\gamma}$) = (0.64$\pm$0.06$\arcsec$,
0.019$\pm$ 0.02, 43$^{+6}_{-4}$).

The angular Einstein radius $\theta_E$ of the SIS model is directly
related to the central velocity dispersion $\sigma_{SIS}$of the stars
in the galaxy by the equation
\begin{equation}
\theta_E = \frac{4\pi\sigma_{SIS}^2}{c^2} \frac{D_{ls}}{D_{os}},
\label{equ:theta}
\end{equation}
where $D_{ls}$ (respectively $D_{os}$) is the angular cosmological distance
between the lens and the source (resp. observer and source). Using
this formula, we find $\sigma_{SIS} \sim$\,169 \kms\ for our best
model.  Because multiple images are observed, the surface mass density
$\Sigma$ inside the Einstein radius equals the critical surface mass
density $\Sigma_c$. This allows us to calculate the amount of mass
$M_E$ inside the Einstein radius $R_E = 1\, h^{-1}$\,kpc. We find
$M_E = 2.16 \times10^{10}\ h^{-1}\ M_{\odot}$. This mass might be overestimated as
the intervening galaxies G1 and G2 are not included in the lens modeling, due to the lack
of sufficient observational constraints (such as the masses of G1 and G2). However,
Keeton \& Zabludoff (\cite{KEEZAB04}) show that this overestimate is unlikely to be larger
than 6\%. This would result in an overestimate of 3\% in the velocity dispersion, i.e., only 5 \kms.

We can compare $\sigma_{SIS}$ to two estimates of the velocity dispersion
based on the SDSS spectrum of \obj. Shen et al. (\cite{SHEN}) measure
the stellar velocity dispersion of the host galaxy, based on the broadening of absorption
features observed its spectrum. They  derive a surprisingly small value, 
$\sigma_* = 96.4 \pm 13.8$ \kms. We note, however, that this value is based on
a low S/N spectrum of the host galaxy, following a principle component
analysis decomposition of the SDSS spectrum. While this is certainly sufficient to 
measure, statistically, velocity dispersions for a large sample of objects, the accuracy reached 
for individual objects remains limited.

An indirect but more robust estimate of
$\sigma_*$ can be found using the relation between the black hole mass
and the galaxy velocity dispersion derived by G\"ultekin et
al. (\cite{GUL}) for all type of host galaxies. Using the black hole
mass estimate $log(M_{BH}) = 7.24\pm0.27$ M$_{\odot}$ derived by
Shen et al. (\cite{SHEN}) from the width of the H$\beta$ emission line of the QSO, 
we obtain $\sigma_*(M_{BH})= 124 \pm 47$ \kms. We note that the width of the QSO H$\beta$ line
measured by Shen et al. (\cite{SHEN}), FWHM(H$\beta$) $\sim$ 2920 $\pm$ 390 \kms, compares 
well with our measurement from the Keck 
spectra: FWHM(H$\beta$) $\sim$ 3090 \kms. The error associated with the former 
estimate is derived by adding
quadratically the intrinsic scatter in the error in the black hole
mass and by propagating the various errors in the formula of
G\"ultekin et al. (\cite{GUL}). The latter estimate is,
within 1$\sigma$, compatible with the lensing velocity dispersion.

\section{Conclusions}

We have conducted a systematic spectroscopic search for QSOs acting as strong
gravitational lenses on background emission-line galaxies. This  has produced a sample of about
14 potential lenses.

We have presented Keck-AO imaging of one of the most likely candidates, \obj, a QSO at $z=0.120$
whose spectrum additionally displays four emission lines at the same redshift, $z=0.640$. 
After PSF subtraction, we found two faint objects, A and B, within a radius of 2\arcsec\ from the QSO. 

Keck LRIS spectroscopy of \obj\, spatially resolves object A and confirms that it is at $z = 0.640$. Traces
of object B may be seen at the expected position, also at $z = 0.640$. We measured the redshift of 
one of the two galaxies seen in the vicinity of the QSO, galaxy G2 at $z = 0.394$.

The position, flux ratio, and shape of objects A and B are fully compatible with 
simple lens models involving a singular isothermal sphere plus external shear. 
Although lower than in most galaxies in SLACS, the velocity dispersion found from the lens models 
is within 1$\sigma$ compatible
with that estimated from the empirical scaling laws established between the mass of the central 
black-hole and that of the QSO host galaxy. 

Given the available observations, \obj\, is the first example of strong gravitational
lensing by a QSO. Full confirmation will require deep HST observations in the optical, where the
emission lines of the background object are prominent. 

Apart from the exotic nature of \obj, the discovery may open a 
new direction in the study of QSOs, providing a powerful test of existing 
empirical scaling laws in QSOs and allowing us to measure the total radial mass profile
of QSO host galaxies with unprecedented accuracy.

\begin{acknowledgements} The authors would like to thank T. Treu, R. Gavazzi, A. Bolton and T. Boroson for 
helpful discussions.
This work is partly supported by the Swiss National Science Foundation (SNSF). 
SGD and AAM acknowledge a  partial support from the US NSF grants AST-0407448 and AST-0909182, and the Ajax Foundation.  We thank the staff of the Keck Observatory for their expert help during the AO and LRIS observations.
DS acknowledges a fellowship from the Alexander von Humboldt Foundation.
\end{acknowledgements}

\end{document}